\documentclass[aps,prb,twocolumn,epsfig,floatfix,twoside,a4paper,showpacs]{revtex4}

\usepackage{epsfig}

\begin{document} 
\input epsf 
\title{A route to explain water anomalies from results 
on an aqueous solution of salt} 
\author{
D.~Corradini, M.~Rovere and P.~Gallo\footnote[1]{Author to whom correspondence
should be addressed;  e-mail: gallop@fis.uniroma3.it}}
\affiliation{Dipartimento di Fisica, Universit\`a ``Roma Tre'' \\ 
Via della Vasca Navale 84, I-00146 Roma, Italy\\}

\begin{abstract}
\noindent In this paper we investigate the possibility to detect
the  hypothesized liquid-liquid critical point of water in 
supercooled aqueous
solutions of salts. Molecular dynamics 
computer simulations are conducted on bulk TIP4P water and on 
an aqueous solution of sodium chloride in TIP4P water, 
with concentration  c = 0.67 mol/kg. 
The liquid-liquid critical point is found 
both in the bulk and in the solution. Its position in 
the thermodynamic plane shifts to higher temperature and lower pressure
for the solution. Comparison with available experimental data
allowed us to produce the phase diagrams of both bulk water and 
the aqueous solution as measurable in experiments. Given the position
of the liquid-liquid critical point in the solution as obtained from our
simulations, the experimental determination 
of the hypothesized liquid-liquid critical point of water 
in aqueous solutions of salts appears possible. 

\end{abstract}




\maketitle

\section{Introduction}\label{intro}

In the past years several theoretical and computer simulation 
studies have hypothesized possible scenarios 
to explain the thermodynamic anomalies of water in the supercooled region.

The {\it critical point scenario} ascribes the  large increase of
thermodynamic response functions to the presence of a liquid-liquid
(LL) phase transition in the supercooled region 
between a low density liquid (LDL) and a high
density liquid (HDL).~\cite{poole92,poole93} 
In this picture the coexistence line
between the two liquids terminates in a liquid-liquid
critical point (LLCP). From the LLCP 
a line of maxima of correlation length, called 
the Widom line, is emanated.~\cite{franzese07}
The lines of extrema of several thermodynamic
response functions, asymptotically merge on the Widom line
upon approaching the critical point from higher temperatures.~\cite{xu05}

The anomalous behavior of water can alternatively be
related to a {\it singularity free scenario}~\cite{sastry} in which
the increase in thermodynamic response functions is due
to local density fluctuations that do not eventually lead to singularities.

Recently another picture was proposed, the {\it critical point free 
scenario}.~\cite{angell08} In this case the HDL-LDL transition is
interpreted as an order-disorder transition of the first order
and no LLCP occurs.

Experimental investigation of the deeply supercooled region of bulk water
is hampered by nucleation.~\cite{debenedetti,angell83,speedy} 
Indications of the existence of LLCP in supercooled
bulk water have been found.~\cite{mishima98,mishima98nat2}
In particular the LLCP was hypothesized to be located at about
$T\simeq 220$~K and $P\simeq 100$~MPa.

The critical point scenario was formulated
by Poole et al.~\cite{poole92} 
as a result of  computer simulations on water with
the ST2 potential.  This interpretation was confirmed by other simulations
on ST2 water,~\cite{poole93,sciortino97,poole05} 
until the very recent paper by Liu et al.~\cite{liu}
The LLCP was also found in water simulated with other
potentials.~\cite{poole93,sciortino97,yamada,paschek05,paschek08,vallauri,brovchenko,tanaka}

Lattice models of water have indicated that 
upon properly tuning the parameters of the models 
one can switch from one scenario to the 
other.~\cite{sastry,kumar08prl,franzese08}

Simulations on spherically symmetric
ramp potential particles, mimicking the properties
of water, show the existence of LLCP for proper choices of the
parameters.~\cite{xu06,yan}

Possible experimental routes for the
observation of the supercooled region where the LLCP is 
supposed to exist are therefore timely. 

In the natural environment water is almost ever found 
as a solvent in mixtures of two or more components
and water can be more easily supercooled in solutions.~\cite{miyata}
Indications of the possible
presence of a LLCP in aqueous solutions have been
found experimentally,~\cite{archer00,archer00pccp,mishima05,mishima07} 
theoretically~\cite{chatterjee}
and in computer simulations.~\cite{corradini08,corradini09} 
Nonetheless the existence and the 
position of LLCP in the thermodynamic plane for 
an aqueous solution has not yet been determined either in 
experiments or in simulations.

Here we present an extensive molecular dynamics (MD) study of 
TIP4P bulk water and of $c=0.67\, mol/kg$  
sodium chloride aqueous solution, NaCl(aq), in TIP4P water.
This study aims to locate the position of the LLCP in the aqueous
solution and to compare it with the LLCP of bulk water in order to see if a 
measure of the LLCP is experimentally feasible in the solution.
We selected this particular concentration because it was found in experiments 
that it is high enough to show deviation from the bulk behavior but
not so high to destroy water anomalies.~\cite{archer00} The chosen 
salt concentration is close to that of seawater 
$c\sim 0.55\, mol/kg$ and it is well above the
physiological concentration which is $c\sim 0.15\, mol/kg$.

The paper is organized as follows. In Sec.~\ref{methods} the simulation
details are given. Results are presented and discussed in Sec.~\ref{res}
an conclusions are drawn in Sec.~\ref{conclusions}.

\section{Methods}\label{methods} 

The TIP4P water potential has been extensively
used for studying supercooled water and it has been recently shown also to
successfully reproduce the phase diagram of all the stable ice 
phases of water.~\cite{sanz}
MD simulations on this potential~\cite{sciortino97} 
estimated the LLCP to be at $T<200$~K and
$P>70$~MPa. However the exact position has not been clearly determined so far.
Ion-ion and ion-water parameters are taken from 
Jensen and Jorgensen.~\cite{jensen}

The two systems are studied upon supercooling and the range of temperatures 
spanned goes from $T=350$~K to $T=190$~K. The range of densities studied is 
$0.83\, g/cm^3 \leq \rho \leq 1.10\, g/cm^3$.
The total number of particles in the simulation box is $N=256$. In the case 
of the solution $N=N_{wat}+N_{Na^+}+N_{Cl^-}$ with $N_{wat}=250$ and 
$N_{Na^+}=N_{Cl^-}=3$. 
The cut-off radius is set to 9.0~\AA~and 
the long-range electrostatic interactions are treated by the Ewald 
summation method. A very recent study proved the importance of
the use of Ewald summations~\cite{liu} in the calculation
of the phase diagram especially close to the
critical region.  
Temperature is controlled with the use of the Berendsen 
thermostat.~\cite{berendsen} The longest runs last up to 30 ns and 
we collected a total of 1848 state points.
The total computational time employed is about six central processing unit (CPU) years 
on a single processor. Simulations were performed 
using the parallel version of the {\small DL\_POLY} package.~\cite{dlpoly}

The temperature of maximum density (TMD) lines have been calculated 
from the minima of the isochores.
The liquid-gas (LG) and the LL
limit of mechanical stability (LMS) lines have been 
calculated considering the points where $(\partial P/\partial \rho)_T=0$.
The position of the LLCP point has been estimated considering
the highest temperature of convergence of the isochores in the 
$P-T$ plane~\cite{poole05,xu06}, which is also the upper bound of
the LL LMS, and also considering 
the  development of a flat region in the isotherms plane and the convergence of
the lines of maxima of the isothermal compressibility $K_T$ and of the constant 
volume specific heat $C_V$.  Generally speaking in fact, 
the existence of a critical point
causes large fluctuations in the region around it. As a consequence above the
critical temperature the thermodinamic response functions show loci of extrema 
that all converge on the Widom line at the critical point~\cite{franzese07}.

\begin{figure}[htbp]
\centerline{\psfig{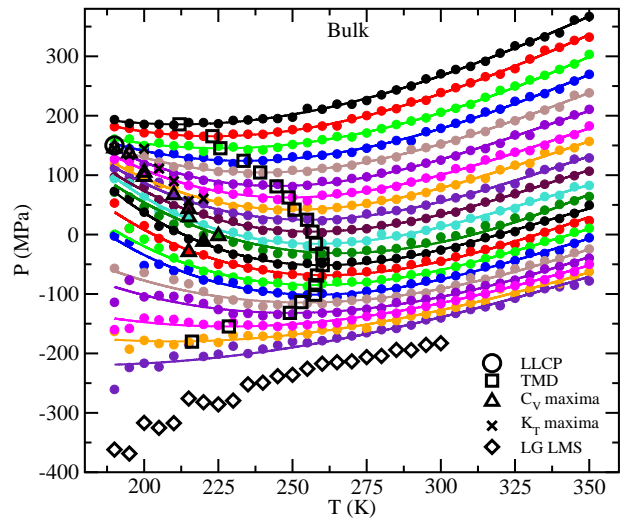}}
\caption{Isochores for TIP4P bulk water in the $P-T$ plane as obtained from
our MD simulations. 
The maxima of $C_V$  and $K_T$  converge to
the LLCP. The TMD points and the LG LMS points are also
reported.
The difference in the isochores density is $\Delta\rho= 0.01\, g/cm^3$, 
the top density is $\rho=1.10\, g/cm^3$ and the bottom density is 
$\rho=0.90\, g/cm^3$. Lines are polynomial fits to the simulated state points.} 
\label{fig:1}
\end{figure}
\begin{figure}[htbp]
\centerline{\psfig{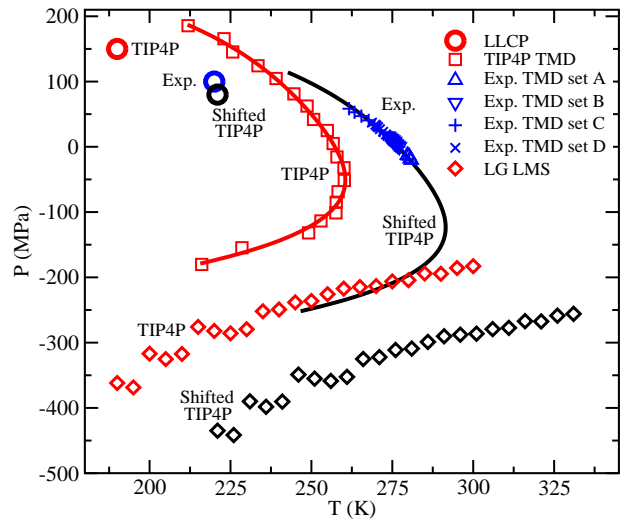}}
\caption{Thermodynamic loci in the $P-T$ plane 
of TIP4P bulk water
as obtained from our simulations and the same loci shifted to match
experimental values.
The match is obtained 
upon applying the shift of $\Delta T=+31$~K and $\Delta P=-73$~MPa. 
We report the position of the LLCP, the TMD line and the LG and 
LMS line.
Sources of experimental values:
LLCP Ref.~\onlinecite{mishima98}, TMD set A Ref.~\onlinecite{henderson}, set B Ref.~\onlinecite{nist}, 
set C Ref.~\onlinecite{harrington}, set D Ref.~\onlinecite{hill}.}
\label{fig:2}
\end{figure}

\section{Results and Discussion}\label{res}

In Fig.~\ref{fig:1} we report for bulk TIP4P water the $P-T$ isochores plane, 
the position of the LLCP of the system, the lines 
of extrema of thermodynamic response functions, 
TMD points and LG LMS line. 
We locate the position of the LLCP for TIP4P 
bulk water to be at $T=190$~K, $P=150$~MPa and $\rho=1.06\, g/cm^3$, being in 
agreement with limits previously estimated for TIP4P bulk 
water.~\cite{sciortino97}

In Fig.~\ref{fig:2} we report the LLCP, the TMD, and 
the LG LMS for the bulk TIP4P
together with the experimental TMD line.
If we now compare the bulk TIP4P TMD and
the LLCP with the experiments we note that upon
shifting the TIP4P phase diagram of $\Delta T=+31$~K and $\Delta P=-73$~MPa,
the TIP4P TMD exactly
superposes to the experimental values and the LLCP falls very close
to its experimentally hypothesized position.
These shifts in pressure and temperatures of the TIP4P phase diagram 
also agree with the shifts found for ice phases of TIP4P.~\cite{sanz}
We also note that the turning point of the shifted TMD line 
falls around $T=290$~K and $P=-120$~MPa which is in the experimentally 
accessible region, above the melting temperature line.~\cite{abascal}
Experiments on stretched
water have reached $P=-200$~MPa.~\cite{green} 

\begin{figure}[htbp]
\centerline{\psfig{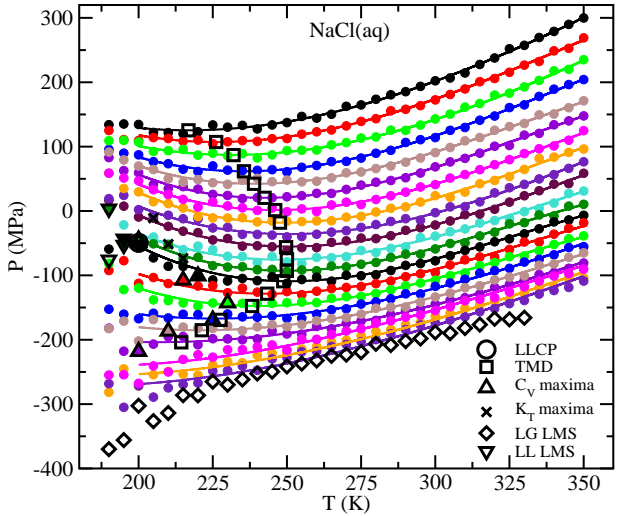}}
\caption{Isochores for $c=0.67\, mol/kg$ NaCl(aq) in the $P-T$ 
plane as obtained from our MD simulations. 
The maxima of $C_V$ 
and $K_T$ converge to the LLCP. 
The TMD points , the LG LMS 
points  and the LL LMS points are also reported.
The difference in the isochores density is $\Delta\rho= 0.01\, g/cm^3$, 
the top density is $\rho=1.10\, g/cm^3$ and the bottom density is 
$\rho=0.90\, g/cm^3$. Lines are polynomial fits to the simulated state points.}
\label{fig:3}
\end{figure}

In Fig.~\ref{fig:3}, similar to the bulk, we show the isochores plane 
for NaCl(aq), the estimated position of the LLCP of the system, the lines 
of extrema of thermodynamic response functions, the TMD points and 
the LL and LG LMS lines. 
For the same set of densities we observe that 
the isochores for the solution globally shift to lower 
pressures with respect to bulk water, with a packing of the isochores 
close to the LG LMS line. 
This shift of the thermodynamic plane has been previously observed in
solutions.~\cite{corradini09,mancinelli}
We locate the position of the LLCP for
NaCl(aq) to be at $T=200$~K, $P=-50$~MPa and $\rho=0.99\, g/cm^3$.
In the solution it was possible to reach equilibrium in our MD simulations
for temperature low enough to determine also the two branches of
the LL LMS line.

\begin{figure}[htbp]
\centerline{\psfig{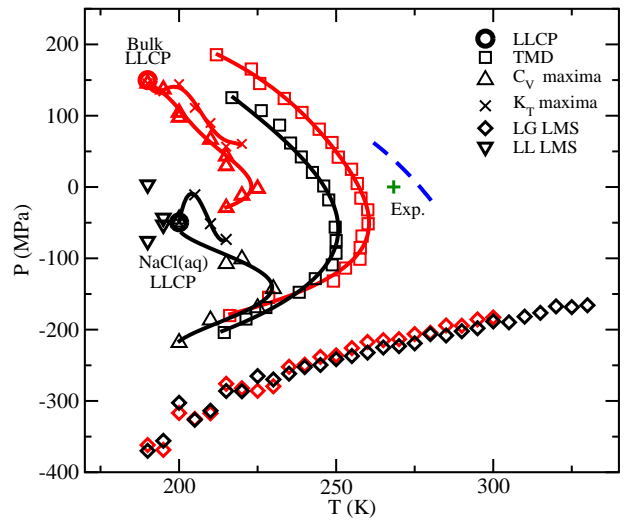}}
\caption{Comparison of the position of the LLCP and of thermodynamic
loci between bulk water (light lines and symbols) and $c = 0.67\, mol/kg$ NaCl(aq) 
(dark lines and symbols) as obtained from our MD simulation. 
The lines are meant as guides for the eye. The dashed line
is the experimental TMD fitted from data reported in Fig.~\ref{fig:2} and the $+$ is
the extrapolated experimental TMD point at 
$P=0.1$~MPa for $c = 0.67\, mol/kg$ NaCl(aq).~\cite{archer00}}
\label{fig:4}
\end{figure}

The results obtained directly from simulations for bulk water and for 
the $c = 0.67\, mol/kg$ NaCl(aq) are summarized in Fig.~\ref{fig:4}.
We also report in the same picture both the experimental TMD line for bulk
water and a TMD point at ambient pressure
extrapolated for our concentration of salt 
from experimental data.~\cite{archer00}
Both in bulk water and in NaCl(aq) the 
lines of maxima of the thermodynamic response functions 
merge in the Widom line very close to the LLCP. 
The LMS line does not move in the solution with respect to the
bulk. The TMD shifts of 10 K to lower temperatures and 
of circa 20 MPa to lower pressures. 
The same shift in temperature is found between the 
experimental TMD line of bulk water and the 
TMD point of the solution, at ambient pressure showing that
our MD is able to capture not only the direction but also the magnitude 
of the shift in temperature found in experiments and confirming
that TIP4P is a very good potential for simulating water. 
We therefore infer that the shift in temperature observed
for the LLCP between bulk TIP4P water and TIP4P water in 
solution is the same as it could be observed in experiments. 
The LLCP moves in the solution of 10 K toward higher temperatures
and shifts in the same direction as TMD in
pressure, but of a much larger amount, $200$~MPa.
This is the signature of the fact that ions 
stabilize the HDL region consistently shrinking in pressure 
the LDL region,
in agreement with the fact the ions seem to be more favorably 
solvated in HDL.~\cite{mishima07,souda} This leads 
to an extension  of the region of existence of the phase where
the local structure of water needs less empty spaces.~\cite{mishima98nat2} 
We observed, see Fig.~\ref{fig:2}, that the 
TIP4P bulk phase diagram is shifted of $\Delta T \simeq -30$~K and 
$\Delta P \simeq 70$~MPa with respect to the experimental result.

\begin{figure}[htbp]
\centerline{\psfig{file=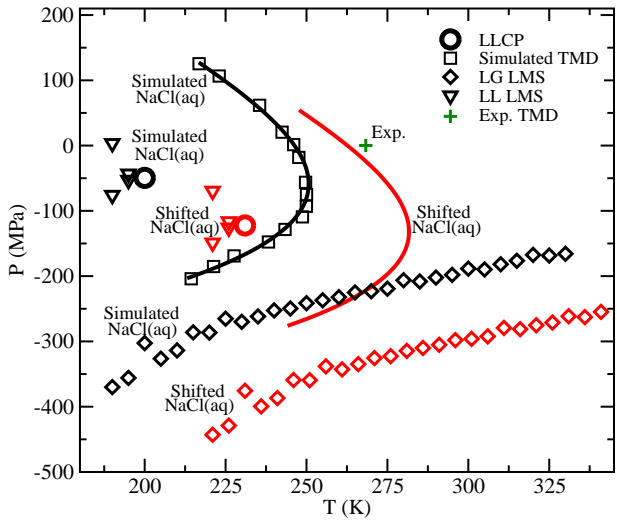}}
\caption{Thermodynamic loci in the $P-T$ plane of $c=0.67\, mol/kg$ NaCl(aq)
as obtained from our simulations and the same loci shifted in pressure and
temperature of the same amount proposed for bulk TIP4P
(see Fig.~\ref{fig:2}), $\Delta T=+31$~K and $\Delta P=-73$~MPa.
We report the position of the LLCP, the TMD line and the LG and LL
LMS lines.
Source of the TMD experimental value: Ref.~\onlinecite{archer00}.}
\label{fig:5}
\end{figure}

On the basis of the comparison with the experimental TMD we hypothesize
that this shift is preserved  for the solution.
We report in Fig.~\ref{fig:5} the phase diagram of NaCl(aq) as obtained
directly from simulations together with the shifted phase diagram.
We observe that the experimental TMD value measured for the solution
falls very close to the TMD shifted curve of the MD simulation.  
We predict for the experimental  $c=0.67\, mol/kg$ NaCl(aq)
a LLCP at around $T_c\simeq 230$~K and
$P_c\simeq -120$~MPa. 
In experiments on water 
large negative pressures  have been obtained, in particular in
NaCl(aq) with concentration $c$=1~mol/kg rupture occurs at $P=-140$~MPa~\cite{green},
therefore the negative critical pressure 
in solution $P_c\simeq -120$~MPa can be reached.

\begin{figure}
\centerline{\psfig{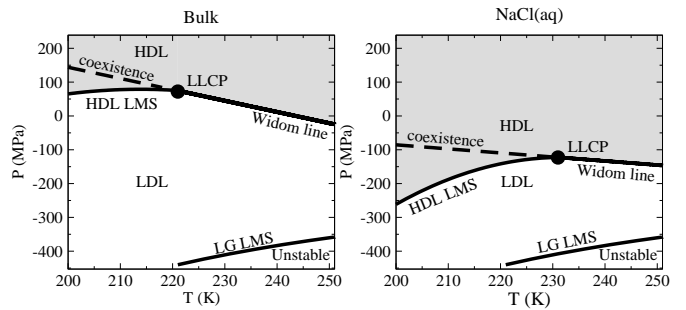}}
\caption{Schematic HDL-LDL phase diagrams obtained from our MD simulations
for bulk water (left panel) and NaCl(aq) (right panel) after applying
the shift needed to match experimental values. 
The LLCP is located at $T\simeq 220$~K and $P\simeq 100$~MPa for bulk water
and at $T\simeq 230$~K and $P\simeq -120$~MPa 
for the $c=0.67\, mol/kg$ NaCl(aq).}
\label{fig:6}
\end{figure}

We report in Fig.~\ref{fig:6} a schematization of the phase diagrams
of both bulk and NaCl(aq) obtained from simulations 
results shifted to  match the experimental values. 
We infer that these should be the phase diagrams of 
water and of a $c=0.67\, mol/kg$ NaCl(aq)
aqueous solution as measurable in experiments.
We can also see 
how the LDL region in the solution  shrinks with respect to the bulk.
On the basis of extrapolation of the data at ambient pressure
for other concentrations of NaCl(aq)~\cite{miyata} 
we can estimate that the homogenous nucleation line for our concentration
should be shifted of circa 5~K downward with
respect to bulk water (see e.g. Fig.~1 in Ref.~\cite{kanno75}). The
LLCP in the solution should then be within the supercooled accessible region
and a measurement of the LLCP appears possible.
Recent experimental studies  on LiCl(aq) have reached 
the supercooled temperature of $T=200$~K~\cite{Mamontov}, 
showing the existence, also in aqueous solutions,
of a fragile to strong dynamical transition typical of
glass formers. In water this dynamical transition happens, 
both in experiments~\cite{chen1} and 
in simulations~\cite{nostrojpcl,xu05}, upon crossing the 
Widom line and therefore appears to be related to the possible
existence of a LLCP. 
The Widom line in water, a network forming liquid, appears 
to play the role of switching line for hopping, 
favored on the side where water is less dense~\cite{nostrojpcl}.

\section{Conclusions}\label{conclusions}

In this paper we have investigated the possibility to detect the hypothesized 
LLCP of supercooled water in aqueous solutions of salts.

We conclude that the phase diagram of
real water is reproducible by TIP4P water upon applying a shift
in temperature and pressure that brings the MD TMD to coincide 
with the experimental values.
The LLCP in this shifted diagram falls at $T = 221$~K and $P= 77$~MPa 
and is located very  close to the value hypothesized 
by Mishima and Stanley,~\cite{mishima98} $T\simeq 220$~K 
and $P\simeq 100$~MPa, corroborating our findings. 

Upon applying an analogous shift to the phase diagram 
of the aqueous solution, 
we predict the LLCP to be experimentally 
detectable  at $T \simeq 230$~K $P\simeq -120$~MPa  
for $c = 0.67\, mol/kg$ NaCl(aq).
The comparison of the phase diagrams of the aqueous solution and
of bulk water shows that the ions stabilize the 
HDL phase consequently shrinking the LDL region.

The existence of a LLCP has been connected to a double well
effective potential between water molecules.~\cite{mishima98nat2}
At low enough temperatures the outer and deep subwell
drives the system to condense in the LDL phase while
the inner subwell at higher enough pressures forces
the transition into the HDL phase. 
If we follow this interpretation our results on the aqueous solution
indicate that the effect of the ions is to shift the inner subwell to higher
distances so that the HDL phase becomes stable at lower pressures 
with respect to bulk water. As a consequence also the TMD line
shifts closer to the LLCP.
The ions instead do not affect the outer subwell since the LG LMS
does not change going from the bulk to the solution. 

Thus we propose as a valuable route for the experimentalists
to solve the issue of the quest of the LLCP of water
to perform measurements on aqueous solution of salts.

\section*{ACKNOWLEDGMENTS}
We thank H. E. Stanley for helpful discussions and for a critical 
reading of the manuscript.
The computational resources of the CINECA
for the ``Progetto Calcolo 891'', of the
INFN RM3-GRID at the
Roma Tre University and of the
Democritos National Simulation Center
at SISSA are gratefully acknowledged.

\end{document}